 \documentclass[preprint2]{aastex}

\slugcomment{Submitted to PASP}

\shorttitle{SARA Remote Observatories}
\shortauthors{Keel et al.}

\begin{document}


\title{The Remote Observatories of the Southeastern Association for Research in Astronomy (SARA)}


\author{William C. Keel\altaffilmark{1}, Terry Oswalt\altaffilmark{2}, Peter Mack\altaffilmark{3},
Gary Henson\altaffilmark{4}, Todd Hillwig\altaffilmark{5}, Daniel Batcheldor\altaffilmark{6},
Robert Berrington\altaffilmark{7}, Chris De Pree\altaffilmark{8}, Dieter Hartmann\altaffilmark{9},
Martha Leake\altaffilmark{10}, Javier Licandro\altaffilmark{11,12}, Brian Murphy\altaffilmark{13},
James Webb\altaffilmark{14}, Matt A. Wood\altaffilmark{15}}


\altaffiltext{1}{Department of Physics and Astronomy, University of Alabama, Box 870324, Tuscaloosa, AL 35487; wkeel@ua.edu}
\altaffiltext{2}{Embry-Riddle Aeronautical University, Daytona Beach, FL; terry.oswalt@erau.edu}
\altaffiltext{3}{Astronomical Consultants and Equipment, Inc., Tucson, AZ; pmack@astronomical.com}
\altaffiltext{4}{East Tennessee State University, Johnson City, TN; hensong@mail.etsu.edu}
\altaffiltext{5}{Valparaiso University, Valparaiso, IN; Todd.Hillwig@valpo.edu}
\altaffiltext{6}{Florida Institute of Technology, Melbourne, FL; dbatcheldor@fit.edu}
\altaffiltext{7}{Ball State University, Muncie, IN; rberring@bsu.edu}
\altaffiltext{8}{Agnes Scott College, Decatur, GA; cdepree@agnesscott.edu}
\altaffiltext{9}{Clemson University, Clemson, SC; hdieter@g.clemson.edu}
\altaffiltext{10}{Valdosta Sate University, Valdosta, GA; mleake@valdosta.edu}
\altaffiltext{11}{Instituto de Astrof\'isica de Canarias (IAC), C/V\'ia L\'actea s/n, 38205 La Laguna, Spain; jlicandr@iac.es}
\altaffiltext{12}{Departamento de Astrof\'isica, Universidad de La Laguna, 38206 La Laguna, Tenerife, Spain}
\altaffiltext{13}{Butler University, Indianapolis, IN; bmurphy@butler.edu}
\altaffiltext{14}{Florida International University, Miami, FL 33199; webbj@fiu.edu}
\altaffiltext{15}{Department of Physics and Astronomy, Texas A\&M University - Commerce, Commerce, TX 75429 ; Matt.Wood@tamuc.edu}

\break

\begin{abstract}
We describe the remote facilities operated by the Southeastern Association for Research in Astronomy (SARA) , a consortium of colleges and universities in the US partnered with Lowell Observatory, the Chilean National Telescope Allocation Committee, and the Instituto de Astrof\'isica de Canarias. SARA observatories comprise a 0.96m telescope at Kitt Peak, Arizona; a 0.6m instrument on Cerro Tololo, Chile; and the 1m Jacobus Kapteyn Telescope at the Roque de los Muchachos, La Palma, Spain. All are operated using standard VNC or Radmin protocols communicating with on-site PCs. Remote operation offers considerable flexibility in scheduling, allowing long-term observational cadences difficult to achieve with classical observing at remote facilities, as well as obvious travel savings. Multiple observers at different locations can share a telescope for training, educational use, or collaborative research programs. Each telescope has a CCD system for optical imaging, using thermoelectric cooling to avoid the need for frequent local service, and a second CCD for offset guiding. The Arizona and Chile instruments also have fiber-fed echelle spectrographs. Switching between imaging and spectroscopy is very rapid, so a night can easily accommodate mixed observing modes. We present some sample observational programs. For the benefit of other groups organizing similar consortia, we describe the operating structure and principles of SARA, as well as some lessons learned from almost 20 years of remote operations.
\end{abstract}



\section{Introduction}

Changes in the instrumental and funding landscapes in astronomy, especially in the USA, have driven increased interest in consortia of universities or other organizations to operate telescopes beyond the reach of any single member, especially as national facilities move toward support of larger telescopes and closing or divestiture of smaller ones. We document here the operation and facilities of one such consortium, the Southeastern Association for Research in Astronomy (SARA). SARA operates three telescopes in the 1-meter class at locations on three continents, using remote internet control. These instruments support a wide range of research, educational, and public-outreach programs.

\section{Sites and telescopes}

\begin{figure*} 
\includegraphics[width=140.mm,angle=270]{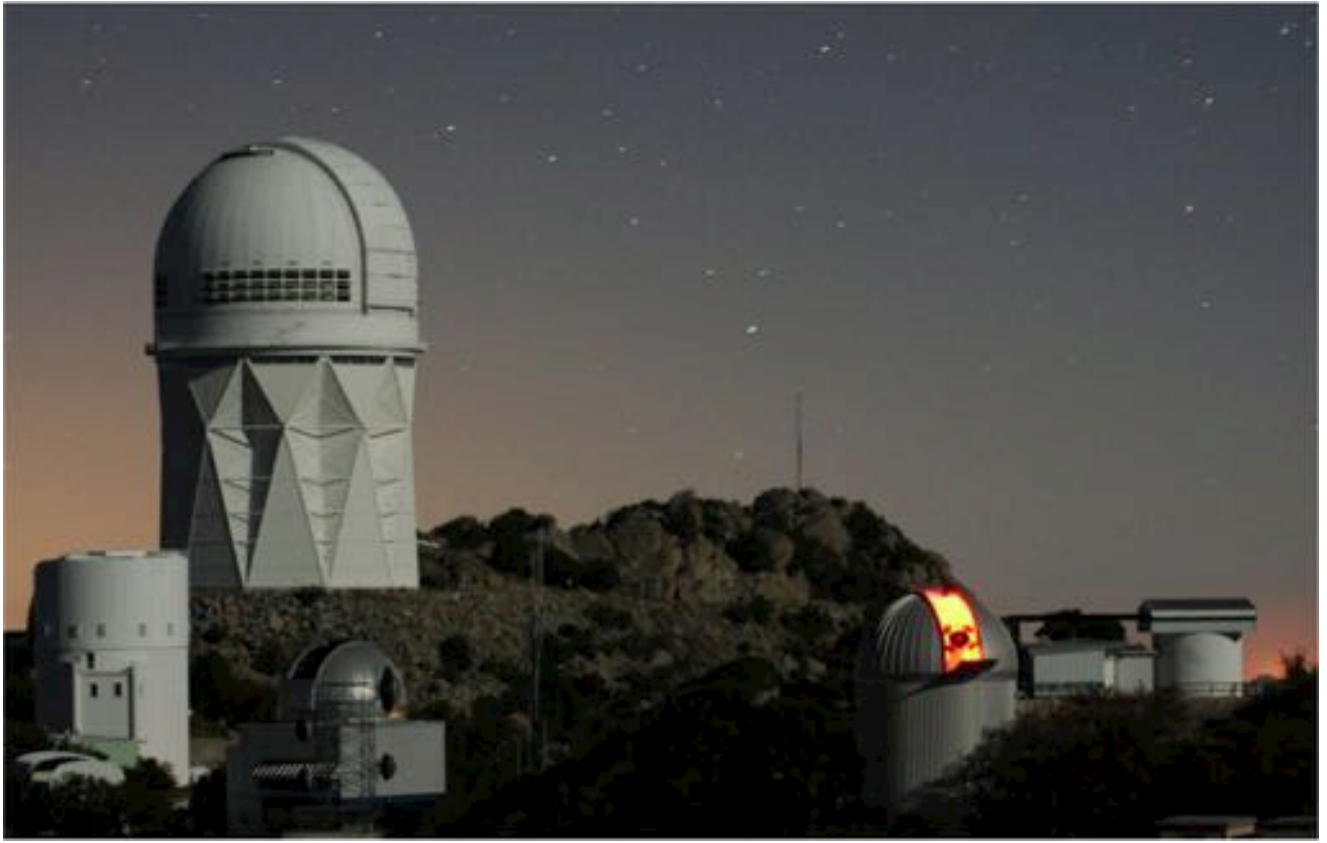} 
\caption{Moonlit view of the SARA-KP site, taken from near the WIYN Observatory. Red dome lights at the SARA telescope were turned on, outlining the open telescope cover petals. (W. Keel)} 
\label{fig-sarakpnight} 
\end{figure*}

The SARA consortium was organized in 1988 in response to an opportunity created by the construction of the 3.5m WIYN telescope at Kitt Peak. To put WIYN at a site with favorable airflow for good image quality, the Kitt Peak \#1 0.9m telescope was removed, and parts from both \#1 and \#2 0.9m telescopes were incorporated into what is now the WIYN 0.9m telescope. The US National
Optical Astronomy Observatories (NOAO) entertained proposals for use of the remaining parts; SARA successfully bid for this, and, starting in 1990, re-created a 0.9m telescope\footnote{As reported by \cite{Glaspey}, the primary mirror had been replaced in 1966 by one about 5 cm larger in diameter, so the SARA telescope at Kitt Peak is more precisely a 1m instrument.} at a site near the Burrell Schmidt telescope above the steep, brush-covered  western slope of Kitt Peak (Fig. \ref{fig-sarakpnight}). Photoelectric photometry was carried out by onsite observers for some years.  By January 1995 it became possible to operate the telescope routinely in remote modes with a CCD camera \citep{Oswalt95}. The quality of long-exposure images is seldom better than 1\farcs 5
FWHM at this site; changing image structure during sequences of short exposures suggests that local atmospheric issues are still a major contributor, though a recent assessment of the mirrors by Nu-Tek Optics showed measurable astigmatism. As other telescopes were added, this was  designated SARA-KP.

\begin{figure*} 
\includegraphics[width=80.mm,angle=270]{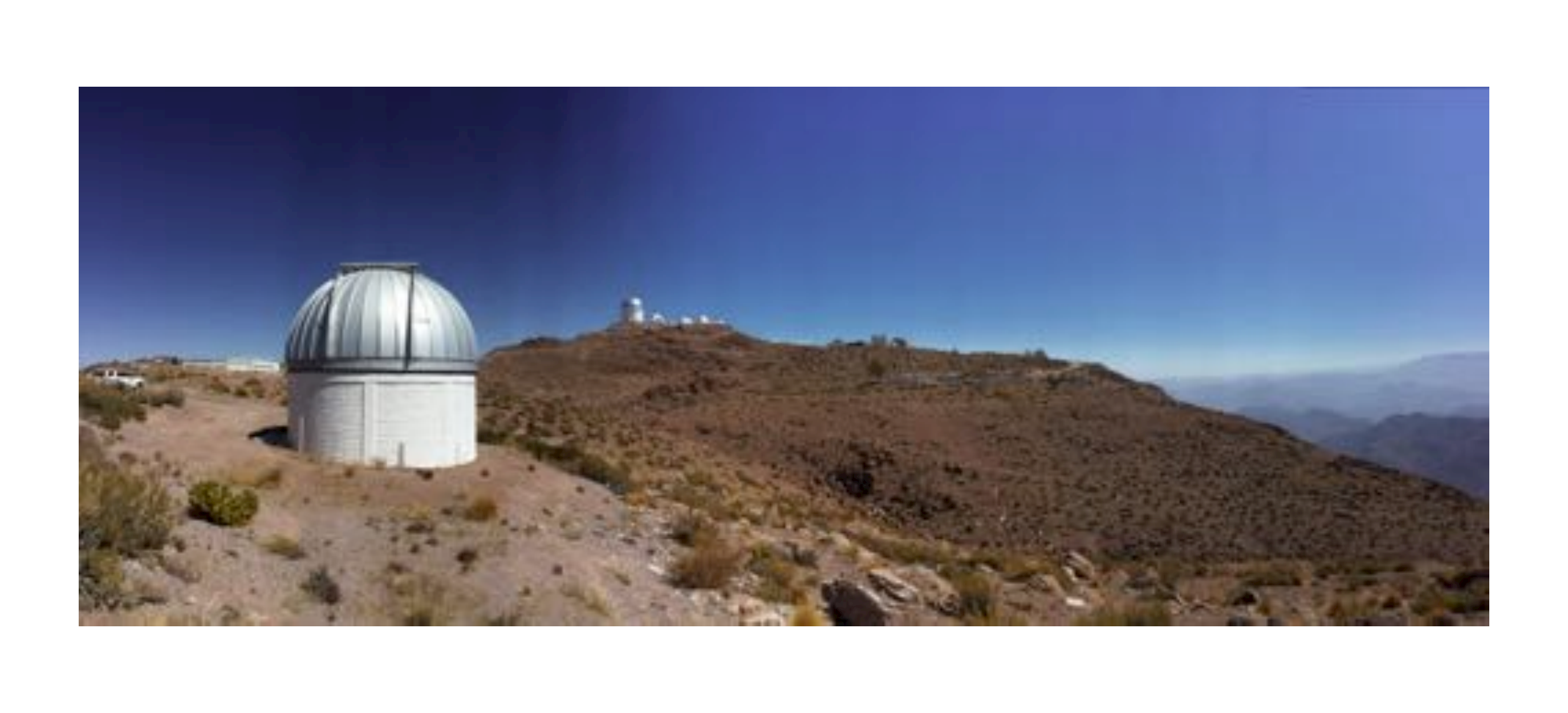} 
\caption{Panoramic view of the SARA-CT installation. The M-EARTH rolloff building is adjacent (to the left), with the Cerro Tololo summit and 4m Blanco telescope in the background. (W. Keel)
} 
\label{fig-saractpan} 
\end{figure*}

Agreements with Lowell Observatory and with NOAO at Cerro Tololo added the 0.6m telescope formerly operated by Lowell at CTIO to the SARA network early in 2010 \citep{Mack2010}; this telescope is now designated SARA-CT. As for other facilities there, the Chilean astronomical community has access to 10\% of the time on this instrument, allocated through the Chilean National Telescope Allocation Committee (CNTAC), and Lowell observers may use a share of telescope time equal to that of the SARA partner institutions. The weather pattern at Cerro Tololo is more favorable than at Kitt Peak, and the seeing is better. The telescope occasionally delivers subarcsecond image quality (mostly in southern summer), and image quality better than 1\farcs 5 FWHM is common. SARA-CT is sited on the southern ridge of Cerro Tololo (Fig. \ref{fig-saractpan}).

\begin{figure*} 
\includegraphics[width=140.mm,angle=270]{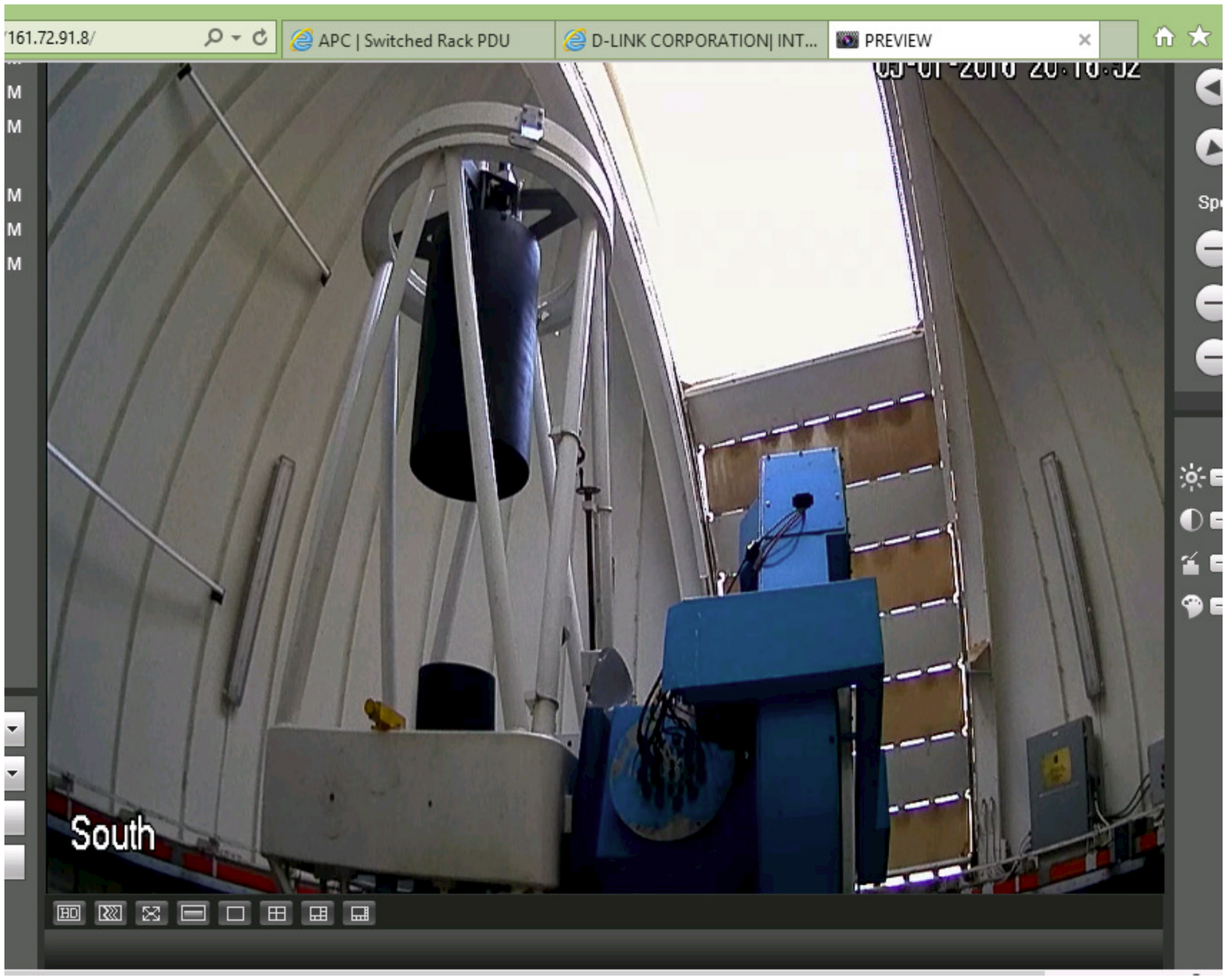} 
\caption{The SARA-RM telescope (former JKT)  as seen with the observer's webcam in June 2016.
} 
\label{fig-sararmwebcam} 
\end{figure*}

The most recent addition to the SARA facilities is the 1m Jacobus Kapteyn Telescope (SARA-RM; Fig. \ref{fig-sararmwebcam}) at the Observatorio del Roque de los Muchachos on the Spanish island of La Palma. Originally opened in 1984, it was mothballed for several years before being refurbished for SARA remote operation by ACE personnel and formally rededicated in late September 2015. The original optical design is shown by \cite{HarmerWynne}; for CCD use with the SARA acquisition assembly, the secondary mirror had to be moved substantially forward (60mm) compared to the original specifications for wide-field photographic plates while retaining the apochromatic correcting lens. This changed the optical properties measurably. The site is excellent and the telescope optics are of high quality; in the months of operation preceding this writing, subarcsecond image quality has been common and values as good as FWHM=0\farcs 5 have been recorded.

All three telescopes are on German-style equatorial mounts, with the telescopes used on the east side of the piers in the north and the west side at Cerro Tololo. The clearance behind the telescopes
is such that reversal of the telescope is not needed to point to any otherwise accessible part of the sky.

The SARA sites are summarized in Table \ref{tbl-sites}. Coordinates at Cerro Tololo are taken from \cite{Mamajek}, who found agreement between GPS and Google Earth coordinates within 3-5 meters. (As Mamajek notes, the SARA-CT site is adjacent to a cluster of markers used for satellite geodetic ranging). Accordingly, we use Google map information to update the latitude and longitude of the other sites (by about 100 meters from that given for the JKT by 
\cite{RGO} , from which we take the elevation listed). The site elevation at Kitt Peak is derived from a USGS topographic map.

\begin{deluxetable}{lccccc}
\tablecaption{Telescopes and Sites\label{tbl-sites}}
\tablewidth{0pt}
\tablehead{
\colhead{Site} & \colhead{Name}   & \colhead{Aperture (m)} & \colhead{Latitude}   &
\colhead{Longitude} &
\colhead{Elevation (m)}  
}
\startdata
Kitt Peak		&	SARA-KP	& 0.96 & 111\degr 35\arcmin 58\farcs 0 W  & +31\degr 59\arcmin 26\farcs 1	&	2073\\
Cerro Tololo	&		SARA-CT & 0.6 & 	70\degr 47\arcmin 57\farcs 11 W  &  -30\degr 10\arcmin 19\farcs 23	& 2012\\
Roque de los Muchachos &	SARA-RM & 1.0 & 	17\degr 52\arcmin 41\farcs 1 W 	 &   +28\degr 45\arcmin 40\farcs 2	&	2369\\
 \enddata
\end{deluxetable}

\section{Instrumentation and remote operations}

The three telescopes have been fitted with updated control hardware, and, when using the same version of the software, the observer is presented with nearly identical interfaces.

The SARA-KP and SARA-CT telescopes have solid tubes, with covers at the top, while the SARA-RM structure uses a Serrurier truss with covers above the primary mirror cell. These covers are computer-operated, with 2 or 4 petals depending on aperture. One such set is visible in Fig 
\ref{fig-sarakpnight}. The opening angle for the petals at Kitt Peak was increased from 90 degrees (out along the tube axis) to about 200 degrees (slightly backwards-facing) after examining the effects of wind shake.

Under a National Science Foundation grant, matching CCD systems from ARC, Inc, of 
San Diego\footnote{www.astro-cam.com} were installed at the Arizona and Chile sites. Each has a $2048 \times 2048$-pixel E2V chip, using closed-cycle cooling to maintain a CCD temperature of -110 C. This is cold enough that dark current ceases to be an important noise contributor even for narrowband imaging with very low sky background. Unattended operation mandates use without regular infusion of cryogens, so the efficiency of  thermoelectric cooling is a key factor. While a white calibration screen was installed at Kitt Peak, most observers find twilight-sky or dark-sky flat fields to be more useful. Tests with multiple exposure times show that the time-dependent illumination patterns due to the bladed shutters of the ARC cameras are reduced well below 1\% for exposures 5 seconds or longer.

An effect of the CCD temperature has been the phenomenon of residual images, from charges trapped by impurities in the chip, and released slowly over a timespan changing with temperature 
\citep{Rest}. The Apogee U42 camera, typically running at -40 C, can release residual charge for several hours from even nonsaturated parts of the image. This can be adequately (if annoyingly) dealt with by taking incremental dark frames between affected targets. Despite using the same CCD architecture, the ARC camera at -110 C shows essentially no release of residual charge even in hour-long exposures. 

Table \ref{tbl-ccds} lists properties of CCD systems used on the telescopes. For completeness it also including imagers employed in the past.

\begin{deluxetable}{lccccl}
\tablecaption{CCD Imager Properties\label{tbl-ccds}}
\tablewidth{0pt}
\tablehead{
\colhead{Site/camera} & \colhead{Pixel scale (\arcsec)}   & \colhead{Field (\arcsec)} & \colhead{Gain}   &
\colhead{Read noise (ADU)} &
\colhead{Dates}  
}
\startdata
SARA-KP ARC	&		0.44	& 899		&	2.3	& 6.0		& 2012-present\\
SARA-KP U42	&		0.38	& 782		&	1.2	& 8.7	& 2006-2012\\
SARA-CT ARC	&		0.38	& 776		&	2.6	& 5.5		& 2013-present\\
SARA-CT E6	&		0.61	& 621		&	1.5	&	5.9	& 2010-2012\\
SARA-CT QSI	&		 0.14	&$343 \times 455$	& 0.46	& 12.4		&2012-2013\\
SARA-CT FLI	&		 0.61	&622		&	2.0	&	9.7 &2015-present\\
SARA-RM Andor Ikon-L &	 0.34	& 697		&	1.0	& 6.3 &	2016-present \\
 \enddata
\end{deluxetable}

\begin{figure*} 
\includegraphics[width=120.mm,angle=0]{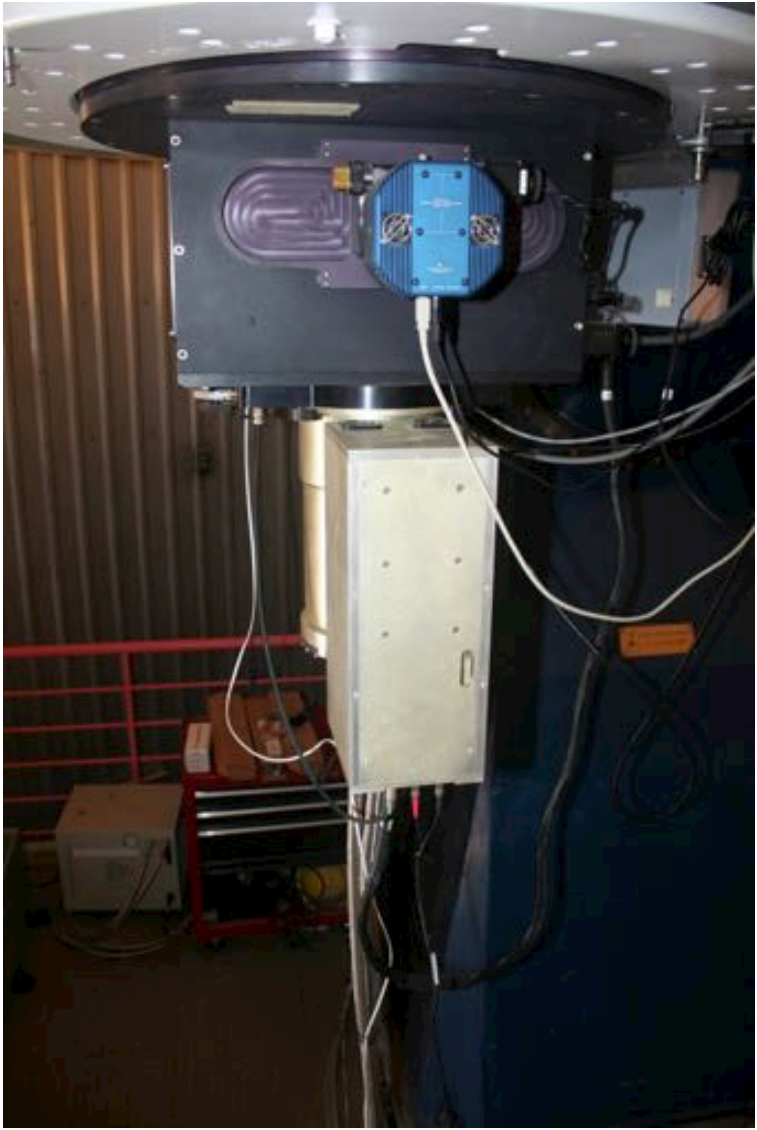} 
\caption{The tailpiece of the SARA-KP telescope in July 2016, showing the acquisition and guiding box above the ARC CCD camera. The covers for the guider track are visible on either side of the octagonal
box housing the heat exchangers for the guide camera's cooling assembly. } 
\label{fig-guider} 
\end{figure*}

\begin{figure*} 
\includegraphics[width=120.mm,angle=270]{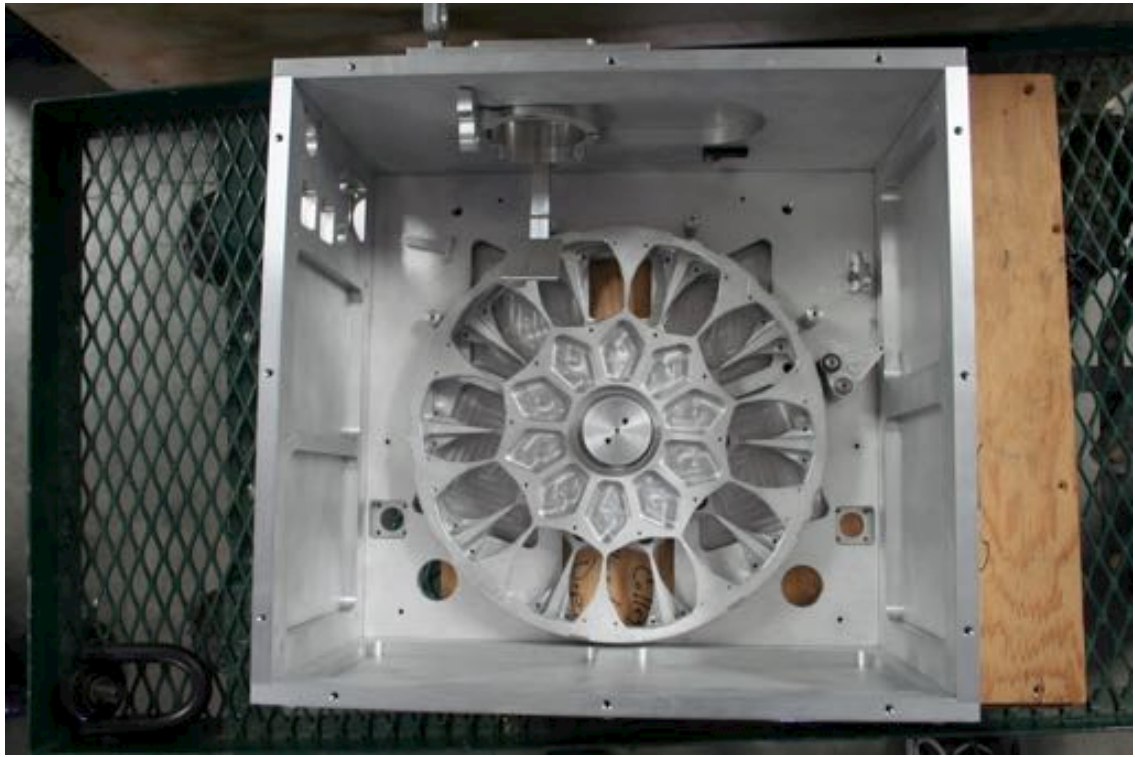} 
\caption{ The acquisition box for SARA-CT being fabricated at the Astronomical Consultants and Equipment, Inc. (ACE) Tucson workshop. The two 10-slot filter wheels are in place but not yet anodized or fitted with fastening hardware. The guider track runs along the wall of the box at the top of the image.} 
\label{fig-filterwheels} 
\end{figure*}

Each telescope has an acquisition/guide box (Fig. \ref{fig-guider}), which includes an autoguider on a moveable single-axis stage, and two filter wheels. The filter wheels (Fig. \ref{fig-filterwheels}) are of different sizes depending on the space available behind each primary mirror. At SARA-KP, the available filters are $UBVRI$ (two sets, one using the Bessel prescription), zero-redshift H$\alpha$, 
H$\beta$ and [O III], redshift-stepped H$\alpha$ with 75-\AA\  FWHM, medium-band continuum including one at 5100 \AA\ also usable for redshifted [O III], neutral-density, and very broad ``white light". At SARA-CT, filters include $UBVRI$, SDSS $ugriz$, white light, zero-redshift H$\alpha$, and a vintage-1975 set of redshift-stepped H$\alpha$ filters now being replaced after suffering degradation with time (especially at the edges). The SARA-RM filter complement is still being filled out, but includes $UBVRI$, $ugriz$, and ``white-light".

\begin{figure} 
\includegraphics[width=80.mm,angle=0]{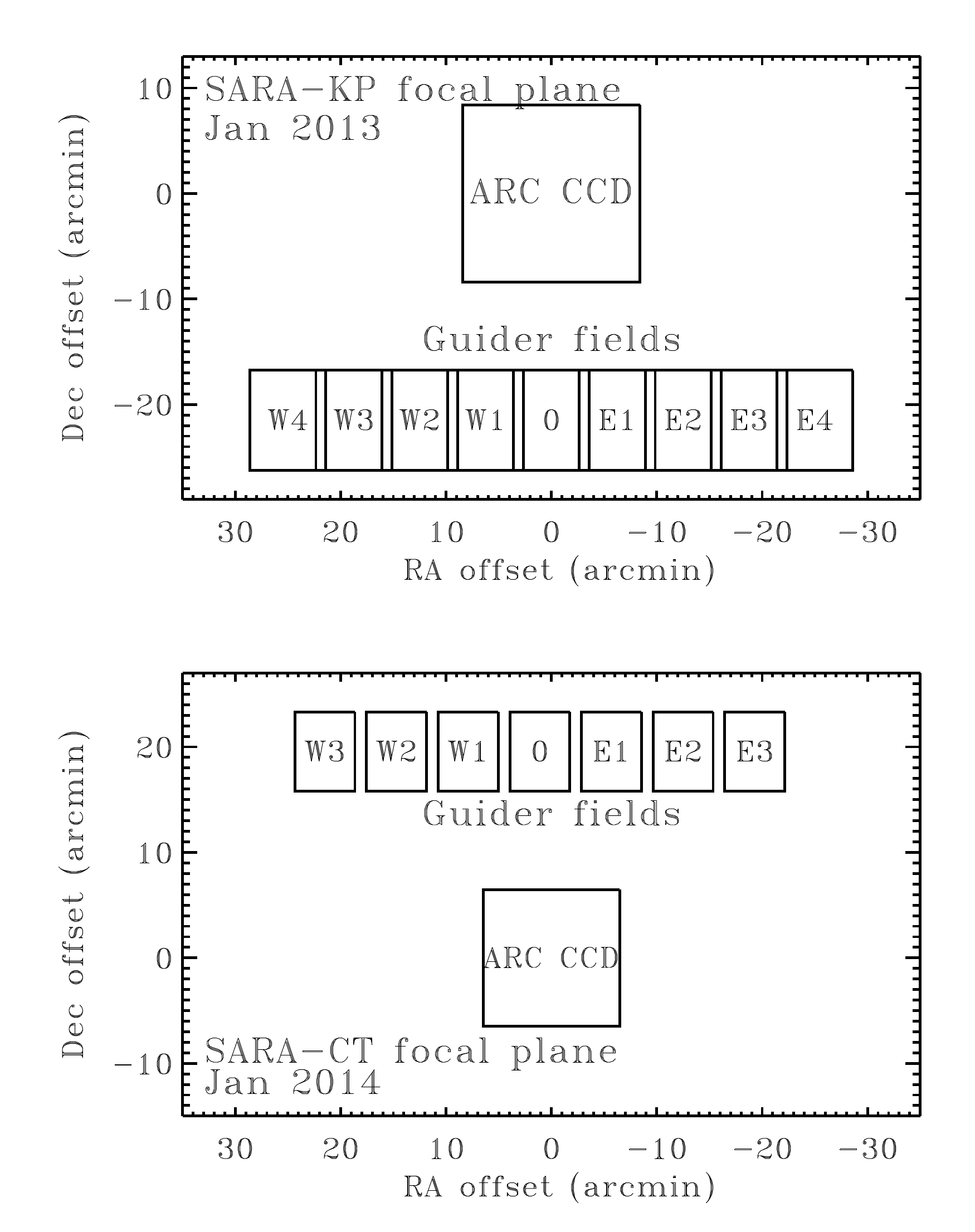} 
\caption{Focal-plane maps for the SARA-KP and SARA-CT instruments, showing the track of the moving guide camera and available offset fields. The E and W designations refer to directions at the back of the telescope, so they are reversed when projected on the sky. At SARA-KP where the location
is well determined, the echelle pickup fiber is nearly centered in the ARC CCD field. The guide
position offsets at SARA-RM have not yet been well-measured on the sky.} 
\label{fig-focalplanes} 
\end{figure}

Each autoguider uses a $2502 \times 3324$-pixel QSI CCD on a one-dimensional movable stage, with focus adjustable to work either with the CCD imager or spectrograph, and to get the best guiding images in view of the large range in distance from the optical axis spanned by the E-W travel of the guider stage. The image scale on the guider is about 0.15\arcsec pixel$^{-1}$. This camera is also available for unfiltered imaging, when parked closest to the optical axis for best image quality; this mode has been used as a fallback when problems occur with the main imaging systems. Guiding uses MaximDL\footnote{From www.cyanogen.com}, feeding correction amplitudes to the telescope control computer. The locations of available guiding fields relative to the ARC imaging CCDs are shown in Fig. \ref{fig-focalplanes}.

In Arizona and Chile, there are nominally identical single-fiber echelle spectrographs, funded by an NSF grant. Achromatic focal reducers give an effective focal length of 3.6m for each telescope, so the fiber aperture maps to 2.8\arcsec (50 $\mu$m diameter, in a 70-$\mu$m cladding), with the fiber on a controllable pickoff assembly. Polymicro FBP fiber is used, giving a transmission for the 20-meter Kitt Peak fiber nominally above 90\% for $\lambda > 4500$ \AA\, falling to 65\% at 3500 \AA\ .  The Finger Lakes CCD covers parts of orders 24-60, with 25-59 uninterrupted and with usable throughput. This chip is operated at typically -45 C; a colder system would improve the limiting sensitivity. The typical resolution is $R$=19,000.

An integrating Astrovid StellaCam camera views the polished jaws of the fiber holder for target acquisition and guiding. Calibration uses a quartz continuum lamp and ThAr comparison source, delivered via fiber with with a lens system matching the focal-reduced telescope beam. The spectrograph fiber assembly can be inserted into the beam in a few seconds; the major time taken for a switch between imager and spectrograph is in changing the focus of both telescope and autoguider.

\begin{figure*} 
\includegraphics[width=140.mm,angle=270]{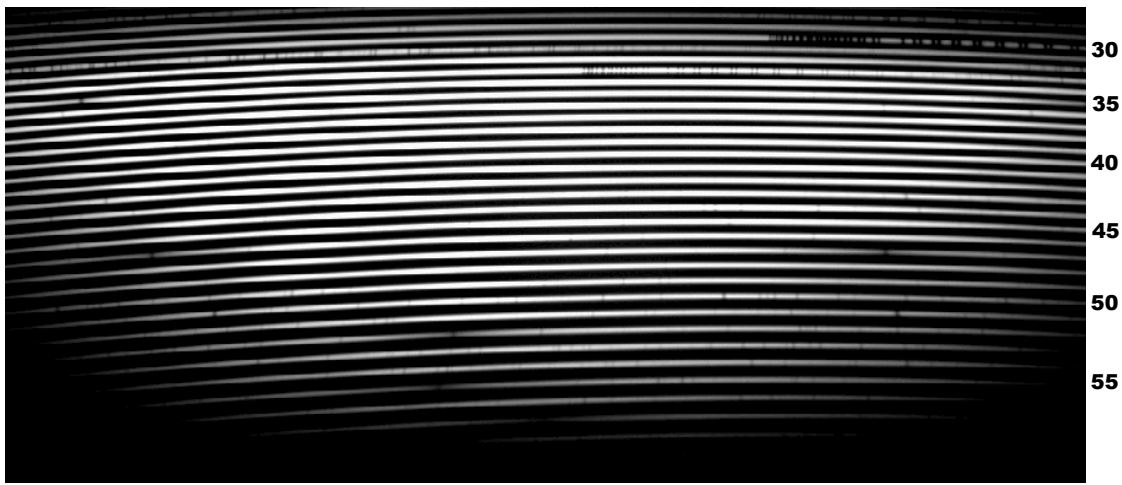} 
\caption{SARA-KP echelle spectrum of Sirius after bias and dark subtraction, showing the layout of spectral orders and location of prominent stellar lines. Order number is indicated on the right, with the edge of the order matching the midpoint of the numerals. The telluric A and B bands of O$_2$ are seen in orders 30 and 33. For display, the pixel scale in the wavelength direction has been compressed by a factor 2. The CCD includes additional area at the bottom, beyond the blue end of useful spectral sensitivity, allowing measurement of scattered light in each exposure. H$\alpha$ (at the low redshifts which are relevant) appears in order 34, with H$\beta$ in orders 45 and 46, and H$\gamma$ in 52.} 
\label{fig-spectrum} 
\end{figure*}

The spectrographs were fabricated by ACE, following a design from Gabor Furesz 
\citep{Mack2013}. The spectral format is illustrated in Fig. \ref{fig-spectrum} using a 10-second exposure of Sirius so the Balmer lines are prominent markers. The exposure level peaks at about 20\% of saturation. A result of the tradeoffs between detector format and spectral format is that the first gap between orders truncates the extreme blue wing of H$\alpha$ for stars with the broadest lines. At the blue end, H$\beta$ and the Ca II K line are each covered by two spectral orders.

\section{Weather monitoring}

At each site the output of an automated weather station is accessible to the control software as well as the observer. There are also all-sky cameras, using fisheye lenses and digital SLR camera bodies. Both of these kinds of information are normally available from other facilities at each site, but a dedicated set increases the chance of data being available on each night in case of data loss from a single source. Dedicated weather stations also provide sensitivity to microclimate (wind and humidity may differ significantly across a single mountaintop). Independently, Boltwood cloud sensors\footnote{Distributed by www.cyanogen.com.} can give a shutdown signal whether or not their associated software is running.

\begin{figure*} 
\includegraphics[width=120.mm,angle=0]{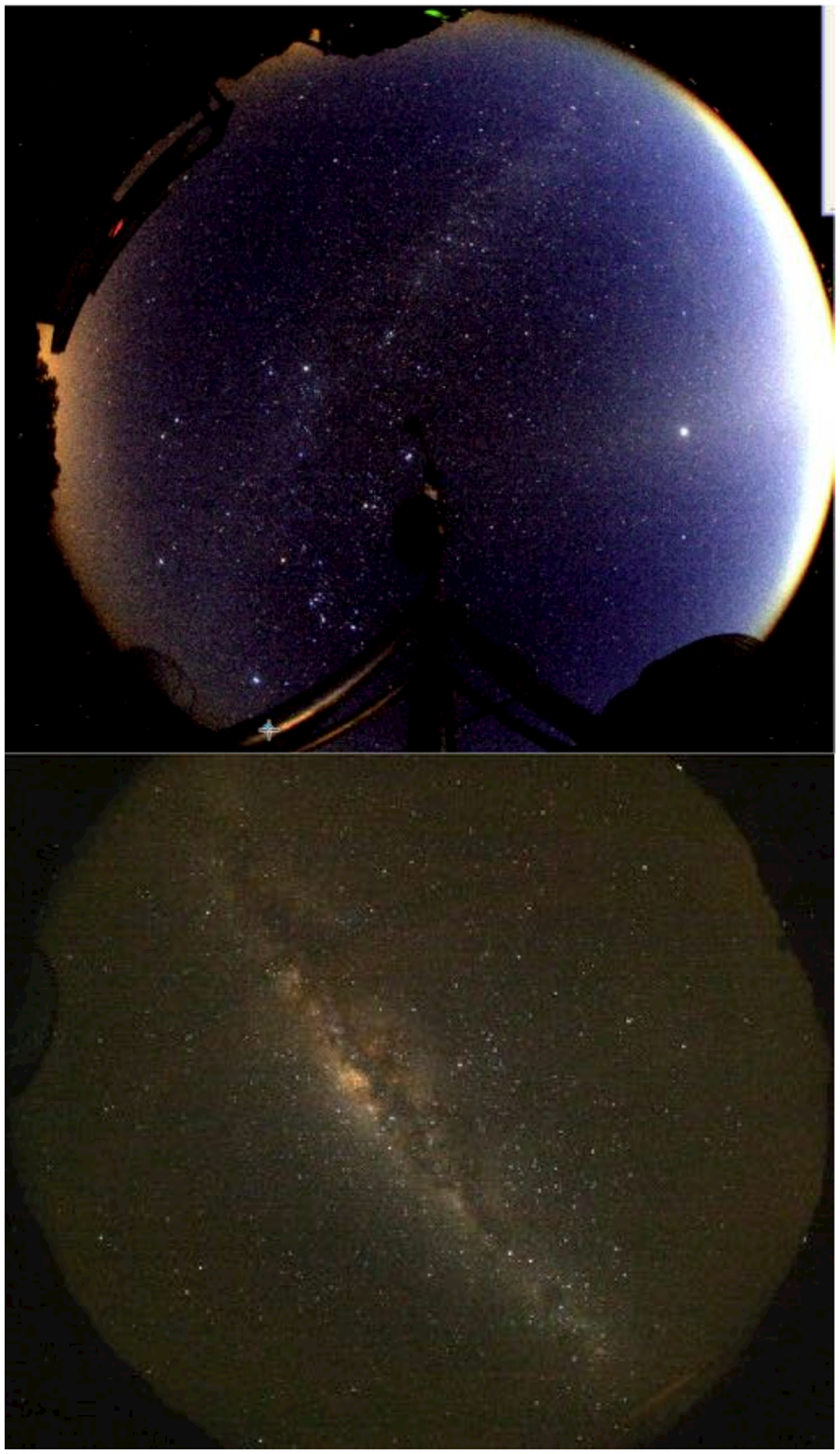} 
\caption{Sample all-sky fisheye images from SARA-KP (top) and SARA-CT. Each is a 30-second exposure with the camera set to ISO 1600. The weather mast at SARA-KP shadows part of the south side of its image.} 
\label{fig-allsky} 
\end{figure*}

The all-sky cameras can operate in greyscale or color modes, and save either JPG or FITS formats if desired. On dark nights, 30-second exposures show detail in the Milky Way and some of the zodiacal band, so clouds can be seen even in dark time (Fig. \ref{fig-allsky}). These systems have seen auxiliary science use, following bright novae and the expanding coma of Comet Holmes up to three months after its October 2007 outburst.

As expected for mountaintop locations, we have lost several anemometers during severe storms.

\section{Telescope control}

\begin{figure*} 
\includegraphics[width=140.mm,angle=270]{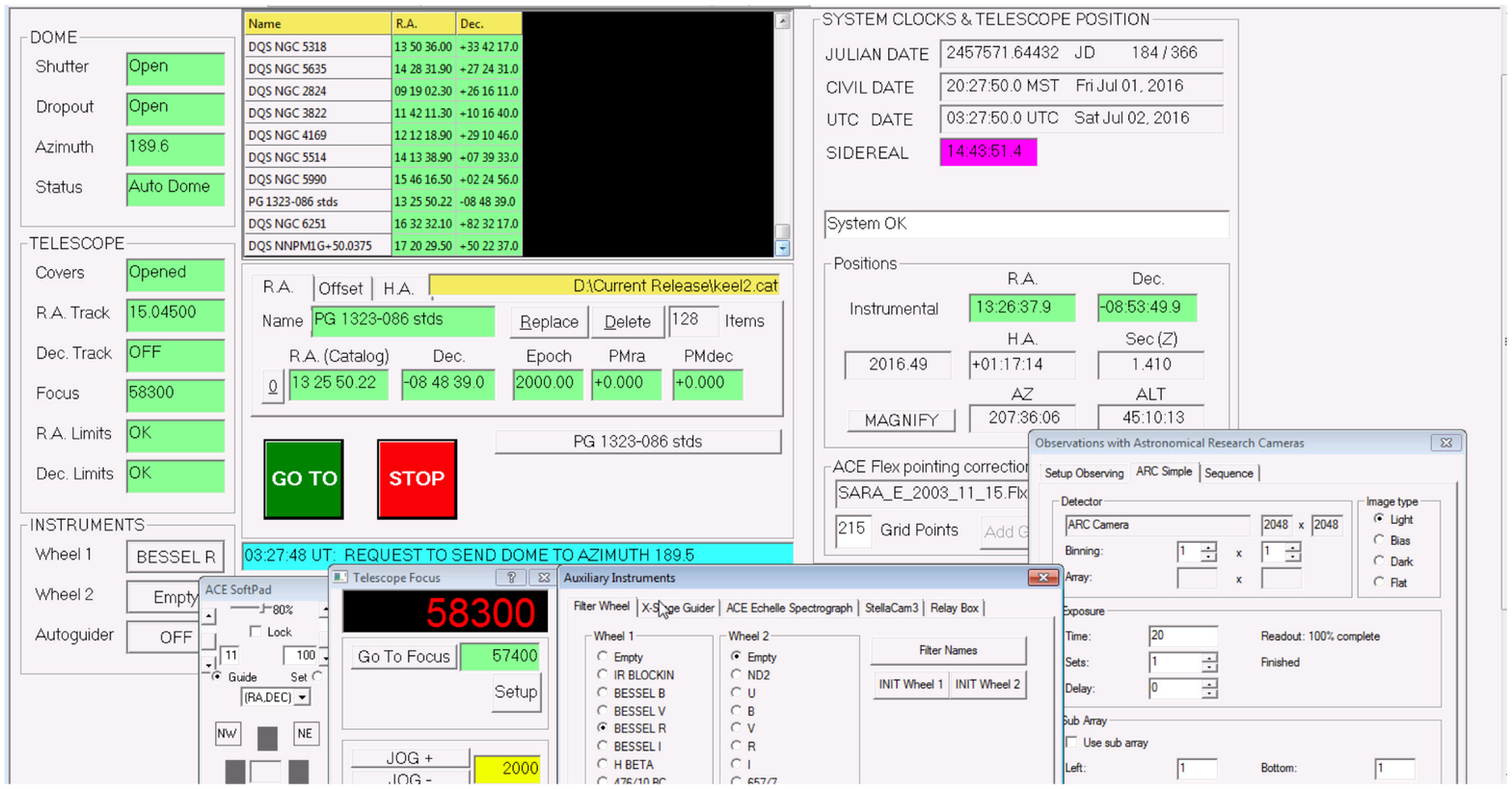} 
\caption{Typical screen view of the telescope-control computer running the ACE control program 
\citep{Mack2011}. Panels for focus, observation parameters, and instrument control float within the overall window. Some panels have tabs for multiple features; for example, the same panel allowing filter selection also includes spectrograph setup and autoguider control.} 
\label{fig-acescreen} 
\end{figure*}

Operation of the observatories is via remote connections to multiple Windows computers at each site, using either Radmin or VNC protocols, through a virtual private network where needed for local security. One machine controls the telescope (Fig. \ref{fig-acescreen}) and passes camera control commands to a second computer which also runs the autoguider and spectrograph detector. A third computer shows the view through an integrating video camera for spectrograph acquisition as well as webcams for general views of telescope and dome. Some functions - weather station, webcams, and CCD control, for example - can be shared by a single PC, so normal use of three at each site provides redundancy in case one computer fails.

Internet switch panels allow power cycling to important systems, including computers (configured to reboot automatically when power comes on) and cameras. For greater safety, conducting brushes deliver power for opening and closing the domes at any orientation. The computers themselves are connected to uninterruptible power supply (UPS) systems.

The remote observer can set filter position, telescope and autoguider focus, and autoguider offset position. These functions also come with hardware initialization positions, in case the absolute encoder value is lost due to system restarts or engineering work. A variety of guider software settings (within MaximDL as well as in the telescope control system) can be set and tuned. SARA-RM uses a more recent version of the ACE control software, which adds integrated exposure series and value fitting for focusing, coordinate retrieval from external name resolvers (such as SIMBAD), planetary ephemerides, and automatic dither patterns for exposure sequences.

Observers can use coordinate catalogs - standard system lists, lists created as objects are observed, or uploaded as text files. These can be sorted or filtered by coordinate, magnitude, or proper motion values.

At SARA-KP there is a control room below the observing floor where all the control computers can be accessed from their consoles; consortium members routinely bring groups of students for training here, and operate the telescope. This room also houses the echelle spectrograph to assist in its temperature control.

Incoming CCD images are displayed in DS9\footnote{ds9.si.edu/site/Home.html}, using 
its XPA messaging function, with MaximDL available for Gaussian image-size measurements and other immediate analysis. A connecting client-server mechanism from ACE handles communication between the PCs running the telescope and CCDs.
	
Software maps of the effective horizon and cable-wrap extent are used to confine pointing directions to safe limits. 

An interruption in internet control does not interrupt observing. This allows not only for occasional communication problems at either end, but an observer can switch locations during an exposure or leave an exposure sequence running and reconnect later.  The systems are designed to close the dome and telescope autonomously when precipitation is detected or weather limits due to humidity or wind speed are violated. For the latter, the observer is queried and must approve continued observing every few minutes.  Closure conditions are: relative humidity 85\% or higher, wind 60 km/hr or above, temperature lower than -10 C, or if the sky gets too bright according to the Boltwood sensor. 
The sunrise condition serves as a fallback for complete loss of internet contact (which happened 
once at Kitt Peak due to problems with work on the line carrying all phone and network traffic to the mountaintop). In any case, it is common (and good practice) for the observer to monitor as well the web compilations of environmental conditions at each site for improved situational awareness. Dome control uses an independent program which can operate even if the ACE control program has been terminated.

Each observer files an online nightly operations report, so usage can be tracked and problems or workarounds communicated. These reports also include recent focus values and pointing or tracking information. These reports are available in an email list as well as a web archive; along with a list for technical discussion, this is important for a distributed organization where successive observers may have no other communication. Important operations announcements, as well as notable images and links to new publications, are disseminated with the Twitter account @SARA\_Obs.

\section{Scheduling}

With minor historical exceptions based on budget, allocation of nights is equal among partner institutions. In practice, each institution submits a list of desired night descriptions. Requests include allowed lunar phases, nights of the week to avoid teaching conflicts, nights needed for scientific or public-outreach reasons, cadence of connected nights, and simultaneous use of multiple telescopes or avoiding simultaneous scheduling. These are then fit into a single schedule. If appropriate, partial nights are easy to arrange, requiring only a change of observer doing the remote controlling. Multiple users can connect at once, allowing new observers to be trained easily. (Consortium policy calls for new observers to share operations with experienced observers for three nights before working solo). Major holiday evenings (Christmas Eve, Christmas, New Year's Eve) can be scheduled on request, but have more stringent weather requirements (no precipitation forecast within 36 hours) because emergency support is not available in the event of, for example, the dome sticking open.

Although operating both a northern telescope and the Cerro Tololo instrument at once requires multiple computers, because of VPN security needs, some observers find it helpful. Subtle variability in blazars on hour timescales was long contentious; a detection is more secure if observed by telescopes thousands of kilometers apart. Simultaneous monitoring can be done in two filters. Finally, some observers favor getting two nights' worth of data at the cost of one night's worth of sleep. Use of SARA-RM and SARA-KP on the same (calendar) night offers the possibility of 18 hours' coverage of northern targets in winter. The two northern sites are separated by 6.2 hours in sidereal time. A much more limited, and seasonally variable, patch of sky is accessible to all three sites at once.

The summer monsoon weather at Kitt Peak motivates an annual closing roughly from July 15 - August 31, when cables are disconnected to reduce risk of damage from nearby lightning strikes. Planned maintenance (mirror realuminizing, for example) is scheduled in this shutdown period. No comparable annual shutdown happens at the other sites.

Typically, engineering nights are scheduled every 3 months during midweek bright time. These allow for planned vacuum pumping of dewars, and resetting the tension on preload balance with seasonal temperature changes. The Kitt Peak telescope in particular is over 50 years old and does have mechanical quirks.

The primary mirrors need cleaning every 1-2 years. The latest cleaning at Kitt Peak improved system throughput by 25\% (and provided a reddening curve for Arizona dust).

\section{Science}

While the astronomers at most member institutions are not numerous, SARA as a whole is essentially a very large virtual astronomy department. As such, the range of scientific studies addressed with SARA facilities is very broad.

In solar-system science, there have been long-running programs to determine asteroid rotation periods, and campaigns to follow short-period comets. Lowell observers and their colleagues have done occultation-related astrometry, for example helping to refine the path of the 29 June 2015 Pluto occultation to place SOFIA near its centerline \citep{Zuluaga}.

In stellar applications, a multiyear study has generated light curves of numerous Mira variables with roughly three samples per month \citep{HensonDeskins}. Several observers use SARA light curves of binary stars for orbital and geometric reconstruction (\citealt{Vaccaro}, \citealt{Samec2013}, \citealt{Samec2015}), and some have identified new variable stars in globular clusters \citep{Murphy}. Hillwig and collaborators (e.g., \citealt{Hillwig} have used SARA data in a long-term effort to identify and characterize binary central stars in planetary nebulae. Several institutions have observers collecting timing data on exoplanet transits, an application which calls for understanding the basis of the computer time stamps as well as when during an exposure the system records the time. SARA was used in a multi-year campaign that discovered the first exoplanet around a post-main-sequence host star, which may have survived engulfment \citep{Silvotti}. Wood and collaborators use the SARA facilities as part of global observing campaigns targeting pulsating white dwarfs or cataclysmic variables (e.g., \citealt{Wood2005}).

Father afield, SARA data have been used to study blazar variations on a range of timescales (e.g., \citealt{Bhatta}, \citealt{Bhatta2016}), variability of Seyfert galaxies, and rapid followup of gamma-ray burst counterparts and tidal-disruption flares in galactic nuclei.  H$\alpha$ images have been used to trace star formation in galaxies, especially in comparison with GALEX and Spitzer data (\citealt{Smith2008}, \citealt{Smith2010}, \citealt{Smith2016}).
After matching resolution to GALEX and XMM Optical Monitor data they have been used to derive optical/UV attenuation curves for backlit galaxies \citep{Keel2014} and for emission-line surveys of AGN with extended ionized clouds \citep{Keel2012}. Some of these projects use the SARA data largely for sample screening, which may not be apparent from publications once the most informative sources have been observed with larger instruments. 

Rapid followup of transient sources (GRB afterglows, recurrent novae, supernovae) is often organized on an ad hoc basis among observers (at least once, \citealt{Swift1644}, beginning with discussion on the cosmoquest.org discussion forum), with results appearing in, for example, more than 50 GCN notices.

With commissioning of the echelle spectrographs, a large atlas of bright-star spectra is being constructed involving observers at FIT and ETSU.

\section{Education and public outreach}

Institutional use of the telescopes provides observational laboratory experiences for both undergraduate and graduate students. This provides access to larger telescopes at higher-quality sites than an institution's local facilities, and in some cases access to otherwise invisible parts of the sky. Some observers use the SARA telescopes as part of public outreach programs, on the campuses or at such external events as Atlanta's DragonCon (where they have supported very well-attended overnight Live Astronomy events since 2007). These events may be organized around a theme (such as star formation and evolution), or driven by audience requests.

Institutions frequently use the more accessible SARA-KP telescope for on-site training of students, either in classes, internal programs or Research Experiences for Undergraduates (REU) programs. A SARA-wide REU program, funded by the National Science Foundation from 1995-2012, included on-site observing and consortium-wide seminars at the beginning and end of each summer session. Many of the results of these projects appeared in the SARA-sponsored journal
JSARA\footnote{jsara.org}, which remains focused on research involving undergraduates. Undergraduates use the instruments frequently for research or class projects; consortium rules require that a faculty observer be present when an undergraduate student moves a telescope.

\section{Consortium Organization and Governance}

Founding institutions of SARA were the Florida Institute of Technology (FIT), Florida International University (FIU), Valdosta State University (VSU), and the University of Georgia (which left in 2007 as faculty retirements changed its departmental research profile). Additional institutions joined as multiple telescopes became available (and needed to be funded).  Table \ref{tbl-members} lists the current member institutions and their accession years. As of 2015, the consortium is headquartered at Embry-Riddle Aeronautical University for administrative and financial matters. In 2015, the Instituto de Astrofisica de Canarias became an associate member. 

\begin{deluxetable}{ll}
\tablecaption{SARA Member Institutions\label{tbl-members}}
\tablewidth{0pt}
\tablehead{
\colhead{Institution} & \colhead{Admission date}   }
\startdata
Florida Institute of Technology		&1989\\
Valdosta State University			&1989\\
University of Georgia 			&1989-2007\\
East Tennessee State University	&1989\\
Florida International University		&1992\\
Clemson University				&1999\\
Ball State University				&2005\\
University of Alabama 			&2006\\
Agnes Scott College				&2006\\
Valparaiso University			&2006\\
Butler University				&2008\\
Texas A\&M University - Commerce	&2012\\
Embry-Riddle Aeronautical University	&2013\\
 \enddata
\end{deluxetable}

The major operating costs for the observatories are the engineering support contract to ACE, and joint-use fees at the host facilities. These are borne by annual dues, currently US\$15,000 per year from each educational partner. This annual budget (slightly under US\$180,000) is very low to operate telescopes on three continents, and we do occasionally experience downtime after major failures, but this model has proven to be sustainable even when many institutions are under financial pressure. This tradeoff allows us to operate the telescopes for less than US\$200 per night, with no additional travel costs.

Each institution appoints a member of the SARA Board of Directors, for three-year terms. The SARA board meets semiannually (in recent years, many of these meetings are conducted online), rotating among member institutions. Formal bylaws call for a consortium chair and board secretary. As additional telescopes have been added, a separate set of facility directors, one per site, manages operational aspects of each instrument. Less formally, a single person has been the telescope scheduler, in a process which has greater impact as the number of partner institutions and telescopes has grown. Institutions with particular dominant science interest can ask to have their allocations weighted toward a particular hemisphere or telescope in each 6-month scheduling period.

The bylaws govern approval of budgetary items (both annual budgets and major expenses arising). As of mid-2016, the consortium is preparing to entertain applications for two or three additional institutional members, now that all three telescopes are in regular operation.

\acknowledgments
Acknowledgements: The ARC imagers and echelle spectrographs for SARA-KP and SARA-CT were funded by the National Science Foundation under grant 0922981 to East Tennessee State University, and the SARA-RM retrofit was funded by the NSF through grant 1337566 to 
Texas A\&M University - Commerce. Ken Rumstay corrected some dating errors in the initial draft; Ron Kaitchuck,Thom Robertson, Carlos Zuluaga, and Stephen Levine provided CCD information.

\facility{SARA}

\appendix
\section{Appendix: Lessons Learned}

Some recurring lessons from SARA remote operations are these, perhaps not all of them obvious.

Apparently robust USB connections may not be.

Remote sites are especially vulnerable to unexpected side effects of operating-system upgrades; we have lost the use of webcams when new OS versions (upgraded in some cases inadvertently) did not have appropriate drivers.

An important start for troubleshooting is a concise listing of what to power cycle, when, and in what order.

Campus network security policies can cause problems. For example, blocking of VNC connections (the only available option for Mac OS or Linux local systems) from floating IP addresses means that laptops cannot be used for class observing (in the most restrictive cases, some observers can work only from off campus).

The SARA all-sky cameras use Canon DSLR bodies and fisheye lenses under clear plastic domes. Degradation of their image quality over several years has been traced to lens coatings gradually turning translucent under constant exposure to sunlight. These have sometimes been the only source of detailed awareness of cloud conditions; even though each site is shared with other facilities hosting all-sky cameras posting to the World-Wide Web, any of these may sometimes undergo outages lasting many nights,

The most current reference source for recent telescope behavior and workarounds has been the online nightly reports by observers. We can scarcely stress strongly enough how much observers should read recent reports before a night's work.

Similarly, for new or occasional observers, the reporting chain for problems (local experienced observer, facility director, only then support engineer) should be documented in an obvious way to avoid unnecessary effort, unwarranted support call-outs, and lost observing time.

The key failure will occur in the one circuit not attached to a remotely-controlled power switch.

Be wary of sending a system just arrived from a vendor to a remote site without doing a full-up test; we have seen some which evidently were never tested between assembly and shipping.

You will always need the larger-capacity uninterruptible power supply.

Judicious tracking of what components to keep spares for, and where they are kept, is essential in reducing downtime at remote locations where shipping times can be long.



\end{document}